\documentclass[10pt]{revtex4}
\usepackage{amssymb}
\usepackage{latexsym}
\usepackage{epsfig}
\usepackage{amsmath}

\begin{document}
\title{New holographic dark energy model inspired by the DGP  braneworld }
\author{A.
Sheykhi,$^{1,2}$,\thanks{asheykhi@shirazu.ac.ir} M. H. Dehghani
$^{1,2}$\thanks{mhd@shirazu.ac.ir} and S. Ghaffari$^{1}$}
\address{$^1$  Physics Department and Biruni Observatory, College of
Sciences, Shiraz University, Shiraz 71454, Iran\\
         $^2$  Research Institute for Astronomy and Astrophysics of Maragha
         (RIAAM), P.O. Box 55134-441, Maragha, Iran}

\begin{abstract}
The energy density of the holographic dark energy is based on the
area law of entropy, and thus any modification of the area law
leads to a modified holographic energy density. Inspired by the
entropy expression associated with the apparent horizon of a
Friedmann-Robertson-Walker (FRW) Universe  in DGP braneworld, we
propose a new model for the holographic dark energy in the
framework of DGP brane cosmology. We investigate the cosmological
consequences of this new model and calculate the equation of state
parameter by choosing the Hubble radius, $L = H^{-1}$, as the system's
IR cutoff. Our study show that, due to the effects of the extra
dimension (bulk), the identification of IR-cutoff with Hubble
radius, can reproduce the present acceleration of the Universe
expansion. This is in contrast to the ordinary holographic dark
energy in standard cosmology which leads to the zero equation of state
parameter in the case of choosing the Hubble radius as system's IR cutoff in
the absence of interaction between dark matter and dark energy.
\end{abstract}
\maketitle
\section{Introduction}
Nowadays there are many strong evidences, on the observational side,
that our Universe is experiencing a phase of accelerated expansion
likely driven by some unknown energy component usually dubbed
``dark energy'' (DE) \cite{Riess}. Recent astronomical
observations indicate that more than $70$ percent of the Universe
consists of DE with negative pressure \cite{Riess}. Disclosing the
nature and origin of such a DE has been one of the biggest
challenges of the modern cosmology. Many theoretical candidates
have been proposed as DE. Among them, those which originate from
fundamental theory such as quantum gravity has arisen a lot of
enthusiasm, recently. For a comprehensive review on DE models see
\cite{Li2011}.

One of the dramatic candidate for DE, that arose a lot of
enthusiasm recently, is the so-called ``holographic dark energy"
(HDE) proposal (see \cite{Hsu,Li, Wang1, Wang2,cop,
Pav`on,Zim,shey0,shey1,shey2} and references therein). This model
is based on the holographic principle which states that the number
of degrees of freedom of a physical system should scale with its
bounding area rather than with its volume \cite{Suss1} and it
should be constrained by an infrared cutoff \cite{Cohen}. It was
shown that in quantum field theory, the UV cutoff $\Lambda$ should
be related to the IR cutoff $L$ due to the limit set of forming a
black hole \cite{Cohen}. If $\rho_D=\Lambda^4$ is the vacuum
energy density caused by UV cutoff, the total energy of size $L$
should not exceed the mass of the system-size black hole
\begin{equation}
E_D\leq E_{BH}\rightarrow L^3\rho_D\leq M_p^2L.
\end{equation}
If the largest cutoff $L$ is taken for saturating this inequality,
we get the energy density of HDE as \cite{Li}
\begin{equation}\label{rho}
\rho_D=\frac{3c^2M_p^2}{L^2}=\frac{3c^2}{8\pi G L^2}.
\end{equation}

The HDE model has been investigated widely in the literature and
has also been tested and constrained by various astronomical
observation \cite{Xin,Feng}. It is fair to claim that simplicity
and reasonability of HDE model provides more reliable frame to
investigate the problem of DE rather than other models proposed in
the literature. It is worth mentioning that in the derivation of
HDE density (\ref{rho}), the black hole entropy $S$ plays a crucial
role \cite{Cohen}. Indeed, the definition and derivation of
holographic energy density depend on the entropy-area relation $S
\sim A \sim L^2$ of black holes, where $A$ represents the area of
the horizon \cite{Cohen}. Any modification of the black holes
entropy due to the quantum correction \cite{Sau} or extra
dimension such as in braneworld scenarios
\cite{Sheykhi1,Sheykhi2} will affect directly on the definition of
the energy density of the HDE and leads to new cosmological
consequences.

The profound connection between thermodynamics and gravity has now
well established through a numerous and complementary theoretical
investigations \cite{Padm}. It has been shown that the
gravitational field equations in a wide range of theories can be
recast as the first law of thermodynamics on the boundary of
spacetime
\cite{Jac,CaiKim,Par,Pad,Cai0,Cai1,Cai2,Cai3,ShHL,CaiHL}. The
studies were also generalized to the brane cosmology, where it was
shown that the differential form of the Friedmann equation on the
brane can be transformed to the first law of thermodynamics on the
apparent horizon \cite{CaiCao,Nozari}. This procedure leads to
extract an expression for the apparent horizon entropy in
braneworld scenarios which is useful in studying black holes
physics on the brane \cite{Sheykhi1,Sheykhi2}. These results
indicate the holographic properties of the gravitational field
equations in a wide range of gravity theories. The deep connection
between the gravitational equation describing the gravity in the
bulk and the first law of thermodynamics on the apparent horizon
reflects some deep ideas of holography.

According to the braneworld scenario, all matter fields in
standard model of particle physics are confined to a brane
embedded in a higher-dimensional bulk, while the gravitational
field, in contrast, is usually considered to live in the bulk
spacetime. A simple and well studied version of braneword model
was proposed by Dvali-Gabadadze-Porrati (DGP)
\cite{DGP,Def1,Def2}. In this model our four-dimensional Universe
is a FRW brane embedded in a five-dimensional Minkowski bulk. It
is important to note that the self-accelerating DGP solution has
ghost instabilities and it is impossible to realize phantom divide
crossing in this branch of solutions without adding extra
component(s). To realize phantom divide crossing in the
self-accelerating branch, it is necessary to add at least a
component as DE on the brane. On the other hand, the normal DGP
branch cannot explain late time cosmic speed-up, but it has the
potential to realize a phantom-like phase by dynamical screening
of the brane cosmological constant. Adding a DE component to the
normal branch solution brings new facilities to explain late time
acceleration and also better matching with observations. These are
the motivations to add DE to this braneworld setup \cite{Nozari}.
In this paper we would like to propose a new modified HDE model in
the context of the DGP braneworld.

The structure of this paper is as follows. In the next section, we
first show that the corresponding Friedmann equation of a flat FRW
Universe in DGP braneworld, can be rewritten as the first law of
thermodynamics on the apparent horizon. This procedure allows us
to derive the entropy expression associated with the apparent
horizon in brane cosmology. Then, having the entropy expression at
hand, we propose a new modified HDE model in DGP braneworld. In
section III, we consider the Hubble radius as IR cutoff and study
cosmological implications of the proposed model. In section IV, we
extend the study to the case where there is an interaction between
the two dark components of the Universe. The last section is
devoted to conclusions and discussions.

%%%%%%%%%%%%%%%%%%%%%%%%%%%%%%%%%%%%%%%%%%%%%%%%%%%%%%%%%%%%%%%%%%%%%%%%%%%%%
\section{The model }\label{model}
We consider a homogeneous and isotropic FRW Universe on the brane
which is described by the line element
\begin{equation}
{\rm d}s^2=-{\rm d}t^2+a^2(t)\left(\frac{{\rm d}r^2}{1-kr^2}+r^2{\rm
d}\Omega^2\right),\label{metric}
\end{equation}
where $k=0,1,-1$ represent a flat, closed and open maximally
symmetric space on the brane, respectively.
Using the spherical symmetry, we can rewrite the metric as
\begin{equation}
ds^{2}=h_{ab}dx^adx^b +\tilde{r}^{2}d\Omega^{2},
\end{equation}
where $\tilde{r}=a(t)r$, $x^0=t$, $x^1=r$ and $h_{ab}$= \rm diag
$\left(-1,\, \frac{a^2}{1-kr^2}\right)$ is the two dimensional
sub-manifold. The dynamical apparent horizon is determined by
equation $h^{ab}\partial_a\tilde{r}\partial_b\tilde{r}=0$, which
implies that the vector $\nabla\tilde{r}$ is null on the apparent
horizon surface \cite{Bak,Hayward}. The explicit evaluation of the
apparent horizon for the FRW Universe gives the apparent horizon
radius
\begin{equation}\label{apparent}
\tilde{r}_A={1}/\sqrt{H^{2}+{k}/{a^2}}.
\end{equation}
where $H=\dot{a}/{a}$ is the Hubble parameter. The associated
temperature $T =|\kappa|/2\pi$ with the apparent horizon is
defined through the surface gravity
\begin{equation}
\kappa=\frac{1}{2\sqrt{-h}}\partial_a\Big(\sqrt{-h}h^{ab}\partial_{b}\tilde{r}\Big),
\end{equation}
The explicit evolution of the surface gravity at the apparent
horizon of FRW Universe reads
\begin{equation}\label{kappa}
\kappa=-\frac{1}{\tilde{r}_A}\bigg(1-\frac{\dot{\tilde{r}}_A}{2H\tilde{r}_A}\bigg).
\end{equation}
Thus the temperature associated  with the apparent horizon is
given by
\begin{equation}\label{tem eq}
T=\frac{1}{2\pi\tilde{r}_A}\bigg(1-\frac{\dot{\tilde{r}}_A}{2H\tilde{r}_A}\bigg).
\end{equation}
where we assume $\dot{\tilde{r}}_A< 2H\tilde{r}_A$. It was shown
that there is indeed a Hawking radiation associated with the
apparent horizon \cite{cao} which gives more solid physical
implication of the temperature associated with the apparent
horizon.

We are going to rewrite the Friedmann equation in the form of the
first law of thermodynamics at apparent horizon. The Friedmann
equation in DGP braneworld has the following form \cite{Def1}
\begin{equation}\label{Friedeq01}
H^2+\frac{k}{a^2}=\left(\sqrt{\frac{\rho}{3M_{\rm
p}^2}+\frac{1}{4r^2_c}}+\frac{\epsilon}{2r_c}\right)^2,
\end{equation}
where $\epsilon=\pm1$ corresponds to the two branches of solutions
\cite{Def1}, and $\rho=\rho_m+\rho_D$ is the total energy density
of the fluid on the brane where $\rho_D$ and $\rho_m$ are energy
density of dark matter (DM) and DE on the brane, respectively. In
the above equation $r_c$ is the crossover length scale between the
small and large distances given by
\begin{equation}
r_c=\frac{M_{\rm p}^2}{2M_5^3}=\frac{G_5}{2G_4}.
\end{equation}
The $\epsilon=+1$ is the self-accelerating solution in which the
Universe may enter an accelerating phase in late time without
presence of additional DE component. The $\epsilon=-1$ branch has
been named as the normal branch, where acceleration only appears
if the DE component is included on the brane. For $r_c\gg1$, the
Friedmann equation in standard cosmology is recovered,
\begin{equation}
H^2+\frac{k}{a^2}=\frac{\rho}{3M_{\rm p}^2}.
\end{equation}
Recent observations indicate that our Universe is spatially flat.
For a flat FRW universe on the brane, Eq. (\ref{Friedeq01})
reduces to
\begin{equation}\label{Friedeq02}
H^2-\frac{\epsilon}{r_c}H=\frac{\rho}{3M_{\rm p}^2}.
\end{equation}
As we mentioned for $\epsilon=+1$ and in the absence of any kind
of energy or matter field on the brane ($\rho=0$), there is a de
Sitter solution for Eq. (\ref{Friedeq02}) with constant Hubble
parameter
\begin{equation}\label{de sitter}
H=\frac{1}{r_c}\Rightarrow a(t)= a_{0} e^{\frac{t}{r_c}},
\end{equation}
which leads to an accelerating Universe with constant equation of
state parameter, $w_D=-1$, exactly like cosmological constant.
However, this solution suffers some unsatisfactory problems. First
of all, it suffers the well-known cosmological constant problems
namely, the fine-tuning and the coincidence problems. Besides, it
leads to a constant $w_D$, while many cosmological evidences,
especially the analysis of the type Ia supernova data indicates
that the time varying DE gives a better fit than a cosmological
constant \cite{Planck1,Planck2,Alam}. In addition, to arrive at
Eq. (\ref{de sitter}) one ignores all parts of energy on the brane
including DE, DM and byronic matter, which is not a reasonable
assumption in a real universe even at the late time.

We also assume that there is no energy exchange between the brane
and the bulk and so the energy conservation equation holds on the
brane
\begin{equation}
\dot{\rho}+3H(1+w)\rho=0,\label{Conserveq}
\end{equation}
while in general there is an interaction between DM and DE on the
brane. Thus both components do not obey the conservation equation,
and they obey instead
\begin{eqnarray}\label{conserveQCDM}
\dot{\rho}_m+3H\rho_m&=&Q, \\
\dot{\rho}_D+3H\rho_D(1+\omega_{D})&=&-Q, \label{conserveQDE}
\end{eqnarray}
where $\omega_D={p_D}/{\rho_D}$ is the equation of state parameter
of DE and $Q=\Gamma\rho_D$ shows the interaction between the DE
and DM on the brane. Using the apparent horizon radius
(\ref{apparent}) we can write the Friedmann equation
(\ref{Friedeq01}) as
\begin{equation}\label{totrho}
\rho_m+\rho_D=3M_{\rm pl}^2\Big[\Big(\frac{1}{\tilde{r}_A}-\frac{\epsilon}{2r_c}\Big)^2-\frac{1}{4{r_c^2}}\Big].
\end{equation}
To obtain a differential form of the Friedmann equation in favor
of the first law of thermodynamics, we take the differential of
equation (\ref{totrho}) and then by using Eqs. (\ref{conserveQCDM}) and (\ref{conserveQDE}) we find
\begin{equation}\label{eq19}
H\rho_D(1+u+\omega_D)dt=2M_{\rm p}^2\Big(\frac{1}{\tilde{r}_A}
-\frac{\epsilon}{2r_c}\Big)\frac{d\tilde{r}_A}{\tilde{r}_A^2},
\end{equation}
where $u={\rho_D}/{\rho_m}$ is the ratio of energy densities of
the two dark components. Multiplying both sides of equation
(\ref{eq19}) by a factor of
$4\pi\tilde{r}_A^3(1-\frac{\dot{\tilde{r}}_A}{2H\tilde{r}_A})$,
and using expression (\ref{kappa}) for surface gravity, we arrive
at
\begin{equation}\label{diffE1}
-8\pi M_{\rm
p}^2\kappa\left(\frac{1}{r_A}-\frac{\epsilon}{2r_c}\right)\tilde{r}_A^2d\tilde{r}_A=4\pi
H\tilde{r}_A^3\rho_D(1+u+\omega_D)dt-2\pi\tilde{r}_A^2\rho_D(1+u+\omega_{D})d\tilde{r}
_A.
\end{equation}
Next, we assume $E=(\rho_m+\rho_D)\frac{4\pi}{3}\tilde{r}_A^3$ is
the total energy inside the $3$-sphere on the brane, where
$V=\frac{4\pi}{3}\tilde{r}_A^3$ is the volume enveloped by
$3$-dimensional sphere. Taking differential form of total energy,
$E$, we find
\begin{equation}
dE=4\pi\tilde{r}_A^2(\rho_m+\rho_D)d\tilde{r}_A+\frac{4\pi}{3}\tilde{r}_A^3(\dot{\rho}_m+\dot{\rho}_D)dt.
\end{equation}
Using Eqs. (\ref{conserveQCDM}) and (\ref{conserveQDE}), we get
\begin{equation}\label{diffE2}
dE=4\pi\tilde{r}_A^2\rho_D(1+u)d\tilde{r}_A-4\pi\tilde{r}_A^3H\rho_D(1+u+\omega_D)dt.
\end{equation}
Combining Eq. (\ref{diffE2}) with (\ref{diffE1}), we obtain
\begin{equation}\label{diffE3}
dE-WdV= 16\pi^2M_{\rm
p}^2T\left(\frac{1}{\tilde{r}_A}-\frac{\epsilon}{2r_c}\right)\tilde{r}_A^2d\tilde{r}_A,
\end{equation}
where we have defined the work density as \cite{Hayward}
\begin{equation}
W=\frac{1}{2}(\rho-p)=\frac{1}{2}\rho_D(1+u-\omega_D).
\end{equation}
The work term $WdV$ is defined as the work done by the change of
the apparent horizon surface. Expression (\ref{diffE3}) is
just the first law of thermodynamics at the apparent
horizon on the brane, namely $dE=TdS+WdV$. We can define the
entropy associated with the apparent horizon on the brane as
\begin{equation}\label{EC eq}
S=\int^{\tilde r_A}_0  dS=16\pi^2M_{\rm p}^2 \int^{\tilde
r_A}_0{\left(\frac{1}{\tilde{r}_A}-\frac{\epsilon}{2r_c}\right)\tilde{r}_A^2d\tilde{r}_A}=\frac{A}{4G_4}\left(1-\frac{\epsilon
\tilde{r}_A}{3 r_c}\right),
\end{equation}
where we have used $M_{\rm pl}^2=(8 \pi G_{4})^{-1}$ and $A= 4\pi
\tilde{r}_A ^2 $ is the area of the apparent horizon. Let us note
that for $\epsilon=-1$, Eq. (\ref{EC eq}) is similar to the result
obtained in \cite{Sheykhi1}. In the limiting case where
$\tilde{r}_A\ll3r_{c}$, one recovers the area law for the apparent
horizon entropy. Physically, this means that the apparent horizon
is not extended in the bulk and located totally on the brane. As a
result, the effect of the extra dimension does not
appear in the entropy expression.\\

It is well-known that in each gravity  theory, the entropy
expression associated with the apparent horizon in cosmology is
the same as the entropy associated with the black hole horizon.
The only change which is needed to replace the horizon radius
$r_{+}$ of black hole with the apparent horizon $\tilde{r}_A$
\cite{Cai2,SMR}. Thus we propose the entropy of the black hole
horizon in the DGP braneworld scenario to be given by
\begin{equation}\label{Sbh}
S=\frac{A}{4G_4}\left(1-\frac{\epsilon r_{+}}{3 r_c}\right),
\end{equation}
where $A=4\pi r_{+}^2$ is the area of the black hole horizon. It
is important to note that the definition and derivation of
holographic energy density ($\rho_{D } =3c^2M^2_p/L^2$) depends on
the entropy-area relationship $S\sim
 A \sim L^2$ of black holes, where $A$ represents
the area of the horizon \cite{Cohen}. Inspired by the entropy
relation (\ref{Sbh}) and following the derivation of HDE
\cite{Gub} and entropy-corrected HDE \cite{Wei,SJ}, we can easily
obtain the corresponding energy density of the HDE in the DGP
braneworld as
\begin{equation}\label{rhoD}
\rho_D=\frac{3c^2M_{\rm p}^2}{L^2}\left(1-\frac{\epsilon
L}{3r_c}\right).
\end{equation}
When $L\ll3r_{c}$, the above equation yields the well-known
holographic energy density, as expected. The significant of the
corrected term in various regions depends on the fraction
$L/r_{c}$. We emphasize here that for studying the HDE in the
framework of DGP braneworld, it is more reasonable to take the
energy density of HDE in the form of (\ref{rhoD}) instead of
(\ref{rho}). This is due to the fact that the well-known area law
for the black holes entropy no longer holds on the brane and the
entropy associated with the horizon on the brane should be
modified as relation (\ref{Sbh}). This is an important point which
was not taken into account in the previous studies on HDE in the
DGP braneworld \cite{HDEDGP}. In the remaining part of this paper,
we shall investigate the cosmological implications of the HDE
density (\ref{rhoD}). Since the simple and natural choice for the
systems's IR cutoff is the Hubble radius $L=H^{-1}$, thus in this
paper we consider this choice in two different cases, namely in
the absence of the interaction and then we consider the
interacting case. In both cases we find the equation of state and
deceleration parameters of a FRW Universe on the brane. We also
plot the related figures to show the evolution of them in each
case.
%%%%%%%%%%%%%%%%%%%%%%%%%%%%%%%%%%%%%%%%%%%%%%%%%%%%%%%%%%%%%%%%%%%%%%%%%%%%%%%%%%%%%%%%%%%%
\section{Non interacting new HDE model} \label{Non in}
Let us consider, for simplicity, the flat FRW universe. The
modified Friedmann equation is given by Eq. (\ref{Friedeq02})
where $\rho=\rho_m+\rho_D$. In the absence of interaction between
DE and DM, the continuity equations read
\begin{eqnarray}\label{ConserveCDM}
&&\dot{\rho}_m+3H\rho_m=0,\\
&&\dot{\rho}_D+3H(1+\omega_D)\rho_D=0.\label{ConserveDE}
\end{eqnarray}
Note that since the matter component is mainly contributed by the
cold DM we ignore the baryon matter component here for simplicity.
From (\ref{Friedeq02}), we can write
\begin{equation}\label{Friedeq04}
\Omega_m+\Omega_D=1-2\epsilon\sqrt{\Omega_{r_c}},
\end{equation}
where we have used the following definitions
\begin{equation}\label{Friedeq05}
\Omega_m=\frac{\rho_m}{3M_p^2H^2},~~~~~~~~~\Omega_D=\frac{\rho_D}{3M_p^2H^2},~~~~~~~~~~\Omega_{r_c}=\frac{1}{4r_c^2H^2}.
\end{equation}
Considering the Hubble radius $L=H^{-1}$ as the system's IR cutoff, we can write
(\ref{rhoD}) as
\begin{equation}\label{rho2}
\rho_D=3c^2M_p^2H^2\left(1-\frac{2\epsilon\sqrt{\Omega_{r_c}}}{3}\right).
\end{equation}
Using (\ref{Friedeq05}), we have
\begin{equation}\label{OmegaDE1}
\Omega_D=c^2\left(1-\frac{2\epsilon\sqrt{\Omega_{r_c}}}{3}\right).
\end{equation}
This equation implies that for HDE in standard cosmology
($\Omega_{r_c}=0$), with Hubble radius as the IR cutoff,
$\Omega_D=c^2$ and thus DE has no evolution during the history of
the Universe. As we know this is not a reasonable result. Thus our
model may resolve this problem, since in our case $\Omega_D$ given
in (\ref{OmegaDE1}) is no longer a constant.

Taking the time derivative of Eq. (\ref{Friedeq02}) and using Eqs.
(\ref{ConserveCDM}), (\ref{ConserveDE}), (\ref{Friedeq04}) and
(\ref{Friedeq05}) we can easily find
\begin{equation}\label{Hdot2}
\frac{\dot{H}}{H^2}=-\frac{3}{2}\left[\frac{1-\Omega_D-2\epsilon\sqrt{\Omega_{r_c}}}
{1-\Omega_D-\epsilon\sqrt{\Omega_{r_c}}-\frac{\epsilon
c^2}{3}\sqrt{\Omega_{r_c}}}\right].
\end{equation}
Taking the time derivative of Eq. (\ref{rho2}), we find
\begin{equation}
\dot{\Omega}_D=\frac{2\epsilon
c^2\sqrt{\Omega_{r_c}}}{3}\frac{\dot{H}}{H}. \label{Omegadot2}
\end{equation}
Using the fact that $\dot{\Omega_D}=H {\Omega}^{\prime}_D$, after
substituting Eq. (\ref{Hdot2}) in (\ref{Omegadot2})  the evolution
of dimensionless ECHDE density as
\begin{equation}\label{Omegadot}
{\Omega}^{\prime}_D=-\epsilon c^2 \sqrt{\Omega_{r_c}}
\left[\frac{1-\Omega_D-2\epsilon\sqrt{\Omega_{r_c}}}
{1-\Omega_D-\epsilon\sqrt{\Omega_{r_c}}-\frac{\epsilon
c^2}{3}\sqrt{\Omega_{r_c}}}\right],
\end{equation}
where the prime denotes derivative with respect to $x=\ln a$.
\begin{figure}[htp]
\begin{center}
\includegraphics[width=8cm]{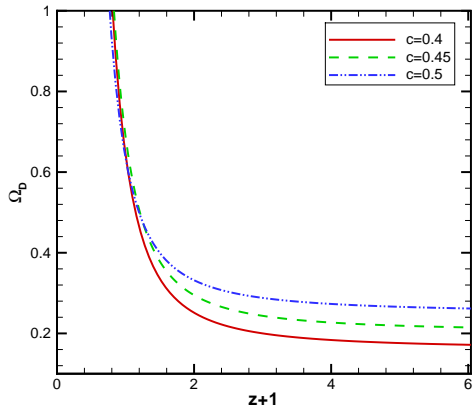}
\caption{The evolution of the EoS parameter $\Omega_D$ versus
redshift parameter $z$.}\label{Omega1}
\end{center}
\end{figure}
Using (\ref{OmegaDE1}) we can  omit $\Omega_{r_c}$ and  rewrite
the above relation in the form
\begin{equation}\label{Omegapr}
{\Omega}^{\prime}_D=-3 (c^2-\Omega_D) \left[\frac{2c^2+\Omega_D
c^2-3 \Omega_D}{c^2+\Omega_Dc^2+c^4-3\Omega_D}\right].
\end{equation}
The evolution of the dimensionless HDE density parameter
$\Omega_D$ as a function of  $1+z=a^{-1}$ is shown in
Fig.(\ref{Omega1}). From this figure we see that at the early
Universe where $z\rightarrow\infty$ we have $\Omega_D\rightarrow
0$, while at the late time, where $z\rightarrow -1$, the DE
dominated, namely $\Omega_D\rightarrow 1$.

Combining Eqs. (\ref{ConserveDE}) and (\ref{rho2}) yields
\begin{equation}\label{EoS2}
\omega_D=-1-\frac{2}{3}\left(1+\frac{\epsilon
c^2\sqrt{\Omega_{r_c}}}{3\Omega_D}\right)\frac{\dot{H}}{H^2},
\end{equation}
\begin{figure}[htp]
\begin{center}
\includegraphics[width=8cm]{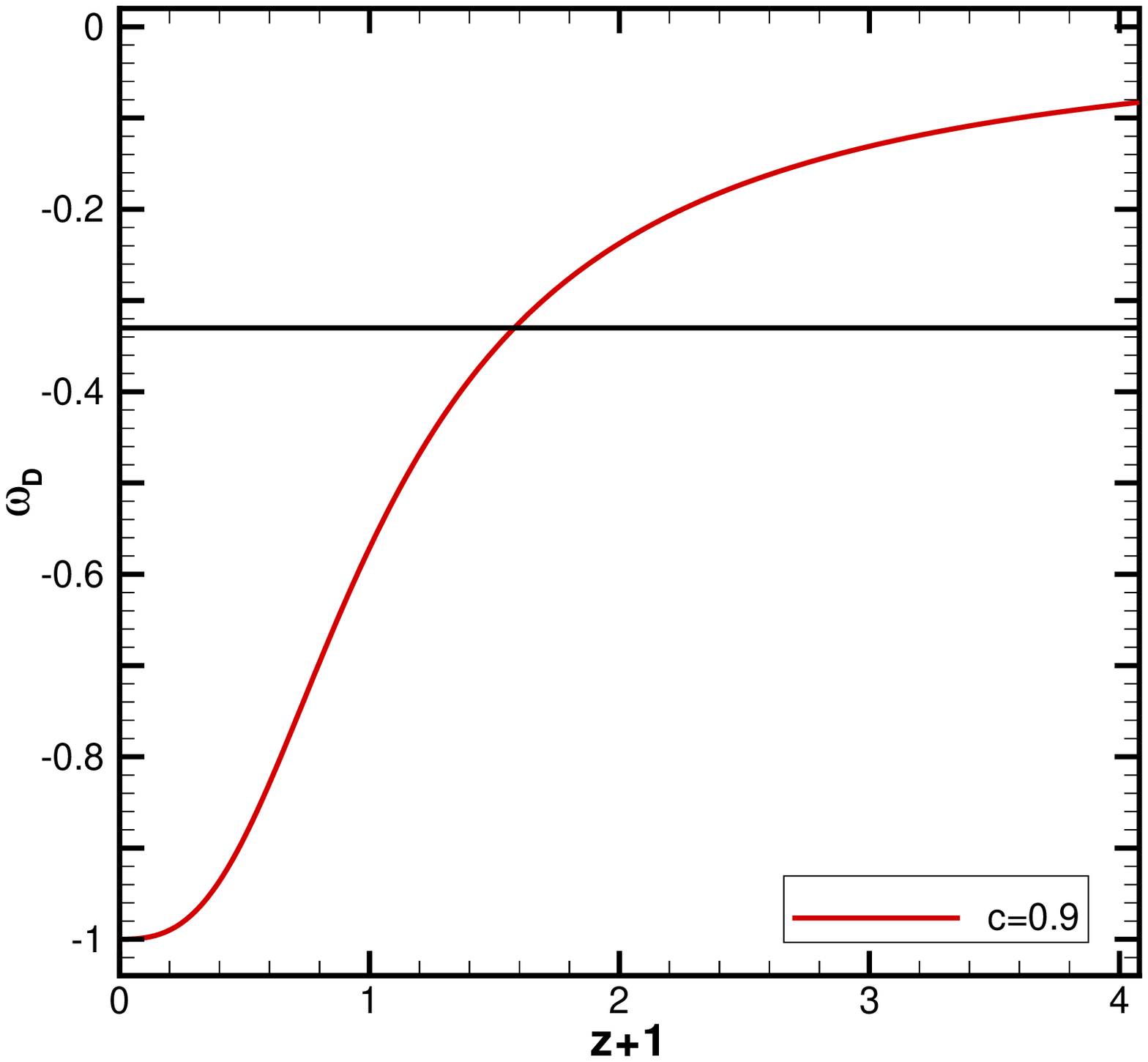}
\includegraphics[width=8cm]{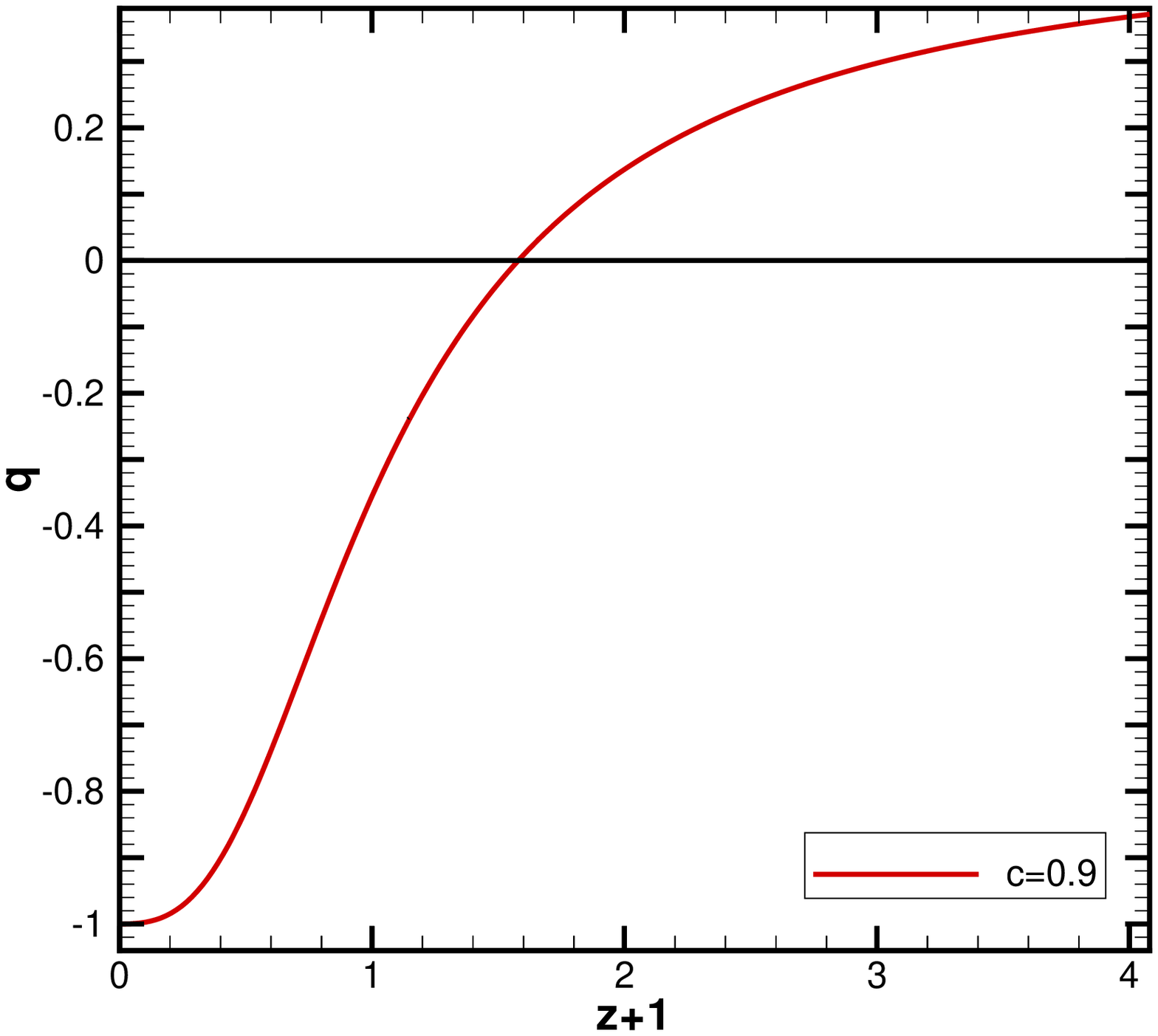}
\caption{The evolution of the EoS and deceleration  versus
redshift parameter $z$ for $c=0.9$.}\label{w1}
\end{center}
\end{figure}
and thus the equation of state parameter is obtained as
\begin{equation}\label{WDE1}
\omega_D=-1+\left(1+\frac{\epsilon
c^2\sqrt{\Omega_{r_c}}}{3\Omega_D}\right
)\left[\frac{1-\Omega_D-2\epsilon\sqrt{\Omega_{r_c}}}
{1-\Omega_D-\epsilon\sqrt{\Omega_{r_c}}-\frac{\epsilon
c^2}{3}\sqrt{\Omega_{r_c}}}\right].
\end{equation}
In the limiting case where $\Omega_{r_c}\rightarrow 0$, we obtain
$\omega_D=0$ which is a wrong equation of state for DE and cannot
derive the acceleration of the Universe expansion \cite{Hsu}. This
is an expected result, since in this regime the effects of the
extra dimension disappear and the standard cosmology is recovered.
It was already shown that in the absence of interaction, choosing
$L=H^{-1}$ cannot produce the cosmic acceleration \cite{Hsu}.
However, as one can see from Eq. (\ref{WDE1}) in the framework of
DGP braneworld with modified HDE density (\ref{rhoD}), the
identification of IR-cutoff with Hubble radius, $L=H^{-1}$, can
lead to accelerated expansion. If we substitute $\Omega_{r_c}$
from  Eq. (\ref{OmegaDE1}), we can further rewrite,
\begin{equation}\label{WDEe3}
\omega_D=\frac{2c^2 (c^2-\Omega_D)}{\Omega_D(c^2+\Omega_D
c^2+c^4-3\Omega_D)}.
\end{equation}
From Eq. (\ref{OmegaDE1}), we see that the limiting case
$\Omega_{r_c}=0$ can be translated to $\Omega_D=c^2$, and thus
from (\ref{WDEe3})  we have $\omega_D=0$. Therefore, the result of
standard cosmology is recovered; no acceleration with Hubble
radius as IR cutoff.

The deceleration parameter is given by
\begin{equation}\label{q}
q=-1-\frac{\dot{H}}{H^2}.
\end{equation}
Substituting Eq. (\ref{Hdot2}) into (\ref{q}) one can obtain the
deceleration parameter as
\begin{equation}\label{qq2}
q=-1+\frac{3}{2}\left[\frac{1-\Omega_D-2\epsilon\sqrt{\Omega_{r_c}}}
{1-\Omega_D-\epsilon\sqrt{\Omega_{r_c}}-\frac{\epsilon
c^2}{3}\sqrt{\Omega_{r_c}}}\right].
\end{equation}
Using relation (\ref{OmegaDE1}) one can rewrite Eq. (\ref{qq2}) in
terms of $c^2$ and $\Omega_D$,
\begin{equation}\label{qq3}
q=-\frac{c^4+6\Omega_D-2c^2\Omega_D-5c^2}{c^2+\Omega_D
c^2+c^4-3\Omega_D}.
\end{equation}
Again for $\Omega_D=c^2$ we have $q=\frac{1}{2}$ which implies a
decelerated phase for the Universe, corresponding to HDE with the
Hubble radius as IR cut-off in standard cosmology \cite{Hsu}. The
evolution of equation of state and deceleration parameters are
shown in figure \ref{w1} for different value of the parameter
$c^2$. From these figures we see that the Universe has a
transition from deceleration to the acceleration phase
$\omega_D<-\frac{1}{3}$. In this model the EoS parameter cannot
cross the phantom line $\omega_D=-1$ in future. In addition the
transition from deceleration to the acceleration phase occurs
around $z\simeq 0.6$ compatible with the recent observations
\cite{Daly,Kom1,Kom2}.

It is worth noting that although the equation of state and the
deceleration parameters do not depend explicitly on the crossover
length scale $r_c$ which is the characterization of the DGP
braneworld, they depend on $r_c$ via the relation between the
dimensionless density parameters in Eq. (\ref{OmegaDE1}). For
$r_c\gg1 (\Omega_{r_c}\rightarrow 0)$, the effects of the extra
dimension vanish and the results of the standard cosmology are
restored \cite{Hsu}.
%%%%%%%%%%%%%%%%%%%%%%%%%%%%%%%%%%%%%%%%%%%%%%%%%%%%%%%%%%%%%%%%%%%%%%%%%%%%%%%%%%%%%%%%%%
\section{Interacting new HDE model} \label{Int}

\begin{figure}[htp]
\begin{center}
\includegraphics[width=8cm]{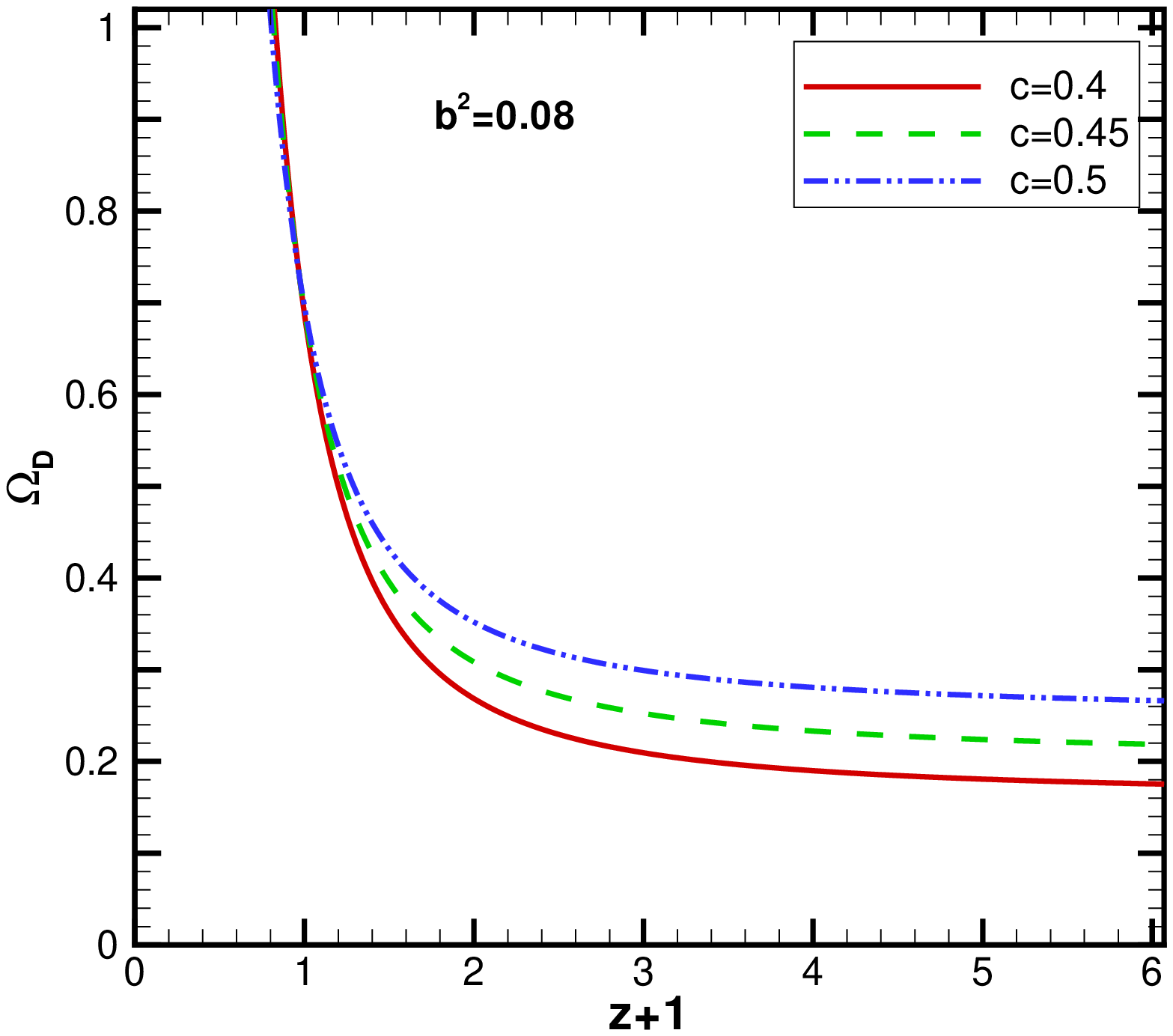}
\includegraphics[width=8cm]{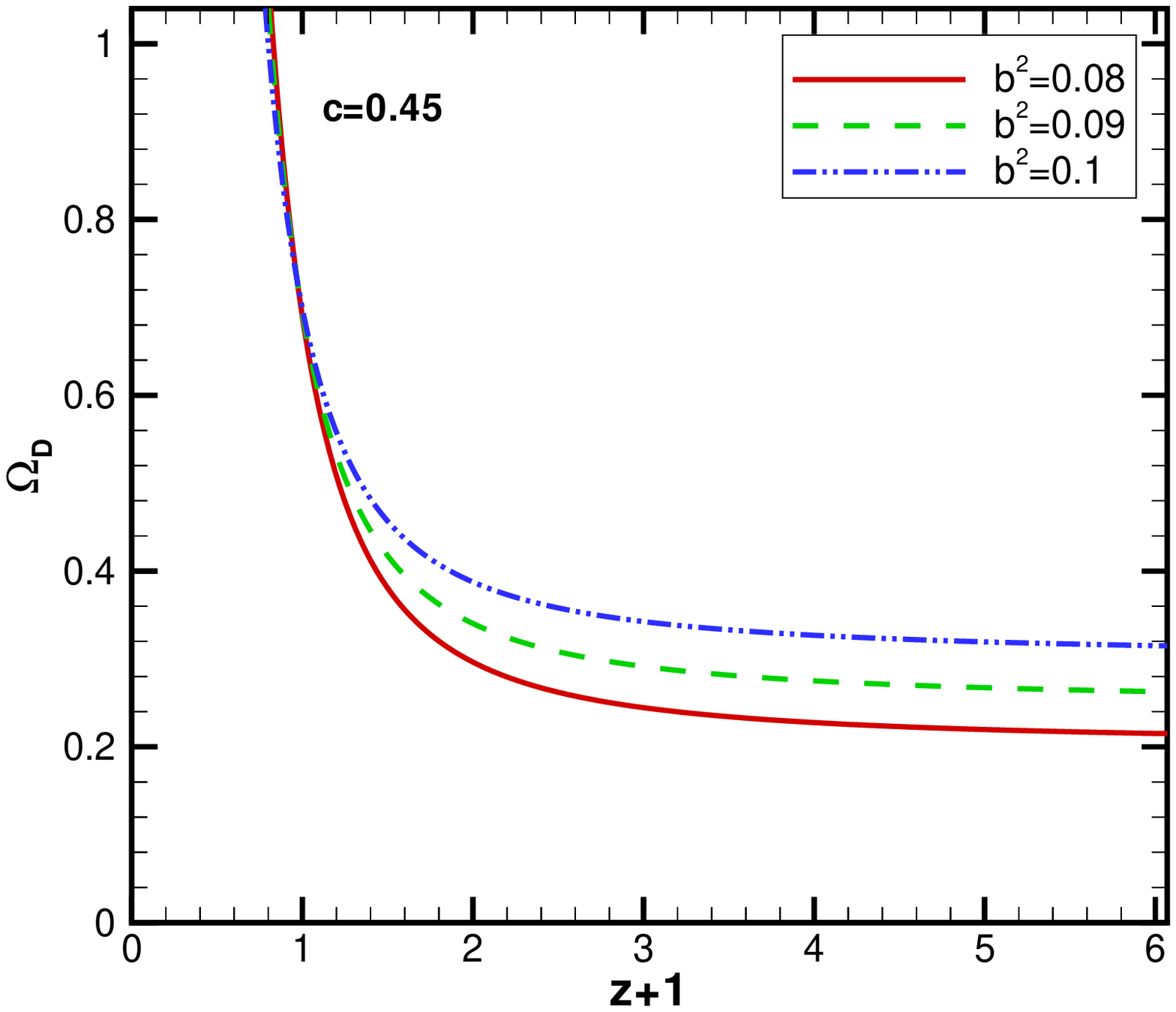}
\caption{The evolution of $\Omega_D$ versus redshift parameter $z$
for interacting new HDE in DGP braneworld.}\label{Omega2}
\end{center}
\end{figure}

In the presence of the interaction between DE and DM, the
conservation equation do no hold separately, they instead obey
(\ref{conserveQCDM}) and (\ref{conserveQDE}). Recent observational
evidences provided by the galaxy cluster Abell A586 supports the
interaction between DE and DM \cite{Bertolami2}. The dynamics of
interacting DE models with different $Q$-classes have been studied
in ample detail in \cite{Amendola}.
Here we choose $Q=3b^2H(\rho_D+\rho_m)$ as the interaction term, where $b^2$ is a coupling constant.\\
Taking the time derivative of Eq. (\ref{Friedeq02}) and using Eqs.
(\ref{Friedeq04}),(\ref{Friedeq05}), (\ref{conserveQCDM})  and
(\ref{conserveQDE}), we find
\begin{equation}\label{Hdot3}
\frac{\dot{H}}{H^2}=\frac{-3(1-\Omega_D-2\epsilon\sqrt{\Omega_{r_c}})+3b^2(1-2\epsilon\sqrt{\Omega_{r_c}})}
{2(1-\Omega_D-\epsilon\sqrt{\Omega_{r_c}}-\frac{c^2\epsilon}{3}\sqrt{\Omega_{r_c}})}.
\end{equation}
Using Eqs. (\ref{Omegadot2}) and (\ref{Hdot3}) we can obtain the evolution of the dimensionless density $\Omega_D$ as
\begin{equation}
{\Omega}^{\prime}_D=3(c^2-\Omega_D)\frac{(-2c^2-c^2\Omega_D+3\Omega_D)+b^2(-2c^2+3\Omega_D)}
{\Big(c^2-3\Omega_D+c^2\Omega_D+c^4\Big)},
\end{equation}
where we have used Eq. (\ref{OmegaDE1}) for omitting
$\Omega_{r_c}$. The evolution of the dimensionless HDE density
parameter $\Omega_D$ as a function of  $1+z=a^{-1}$ is shown in
Fig. (\ref{Omega2}). Again we see that at the early time
$\Omega_D\rightarrow 0$, while at the late time,
$\Omega_D\rightarrow 1$.

Combining Eqs.  (\ref{conserveQDE}), (\ref{Friedeq04}) and
(\ref{Friedeq05}) we arrive at
\begin{equation}\label{EoS3}
\omega_D=-1-\frac{2}{3H}\left(\frac{\dot{H}}{H^2}\right)-\frac{2c^2\epsilon\sqrt{\Omega_{r_c}}}{9\Omega_D}\frac{\dot{H}}{H^2}
-\frac{b^2(1-2c^2\epsilon\sqrt{\Omega_{r_c}})}{\Omega_D}.
\end{equation}
Substituting Eq. (\ref{Hdot3}) into Eq. (\ref{EoS3}), we find the
equation of state parameter of interacting new HDE as
\begin{equation}\label{QWDE3}
\omega_D=\frac{\sqrt{\Omega_{r_c}}(-\epsilon\Omega_D+\frac{\epsilon
c^2}{3}-\frac{2\epsilon^2c^2}{3}\sqrt{\Omega_{r_c}})-{b^2(1-2\epsilon\sqrt{\Omega_{r_c}})(1-\epsilon\sqrt{\Omega_{r_c}})}}
{\Omega_D(1-\epsilon\sqrt{\Omega_{r_c}}-\Omega_D-\frac{c^2\epsilon}{3}\sqrt{\Omega_{r_c}})}.
\end{equation}
Substituting Eq. (\ref{Hdot3}) into (\ref{q}) we obtain the
deceleration parameter as
\begin{equation}\label{Qq3}
q=\frac{1-4\epsilon\sqrt{\Omega_{r_c}}-\Omega_D+\frac{2c^2\epsilon}{3}\sqrt{\Omega_{r_c}}+3b^2(1-2\epsilon\sqrt{\Omega_{r_c}})}
{2(1-\epsilon\sqrt{\Omega_{r_c}}-\Omega_D-\frac{c^2\epsilon}{3}\sqrt{\Omega_{r_c}})}.
\end{equation}
\begin{figure}[htp]
\begin{center}
\includegraphics[width=8cm]{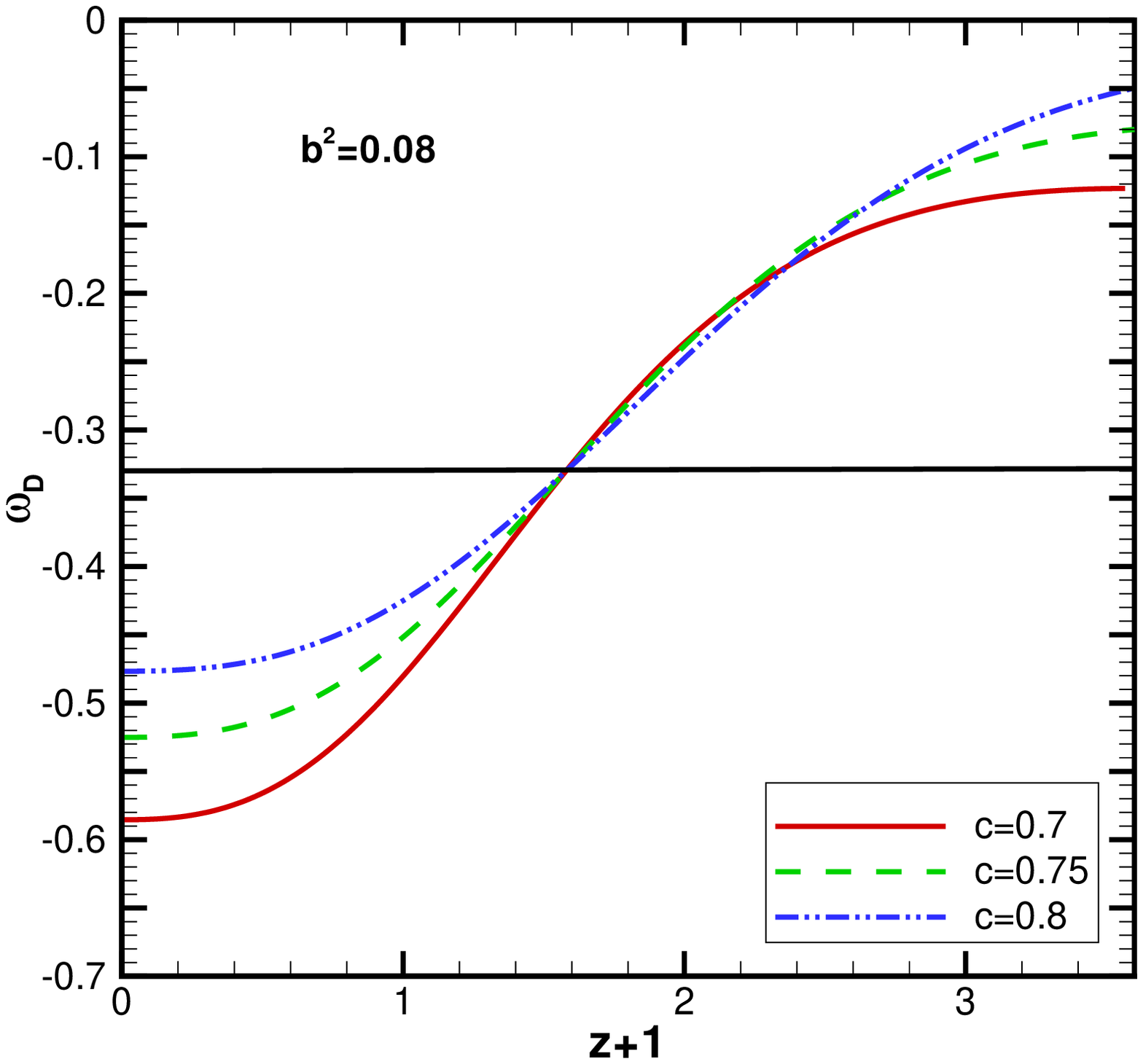}
\includegraphics[width=8cm]{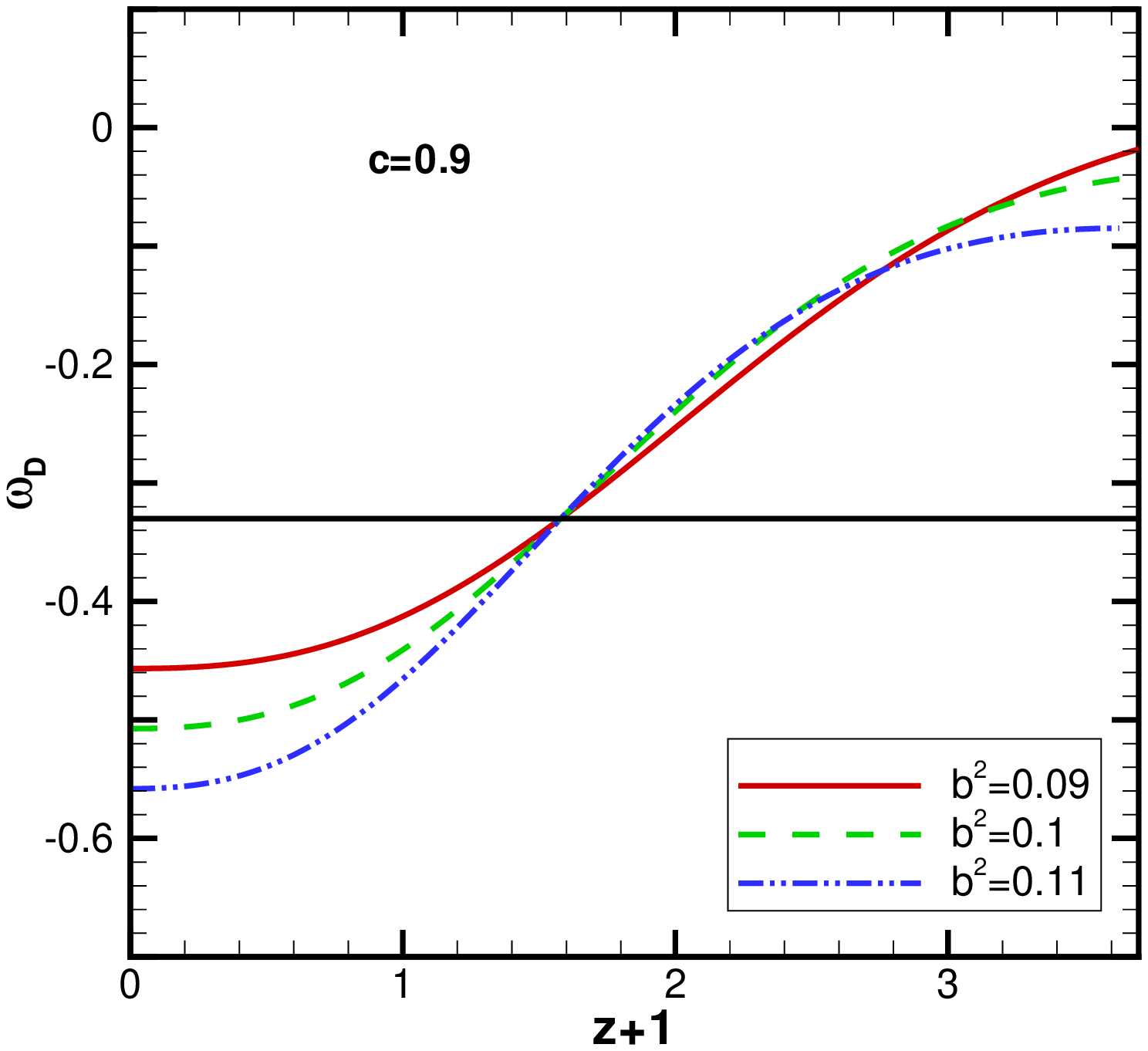}
\caption{The evolution of the EoS parameter $\omega_D$ versus
redshift parameter $z$ for interacting new HDE in DGP
braneworld.}\label{w2}
\end{center}
\end{figure}
\begin{figure}[htp]
\begin{center}
\includegraphics[width=8cm]{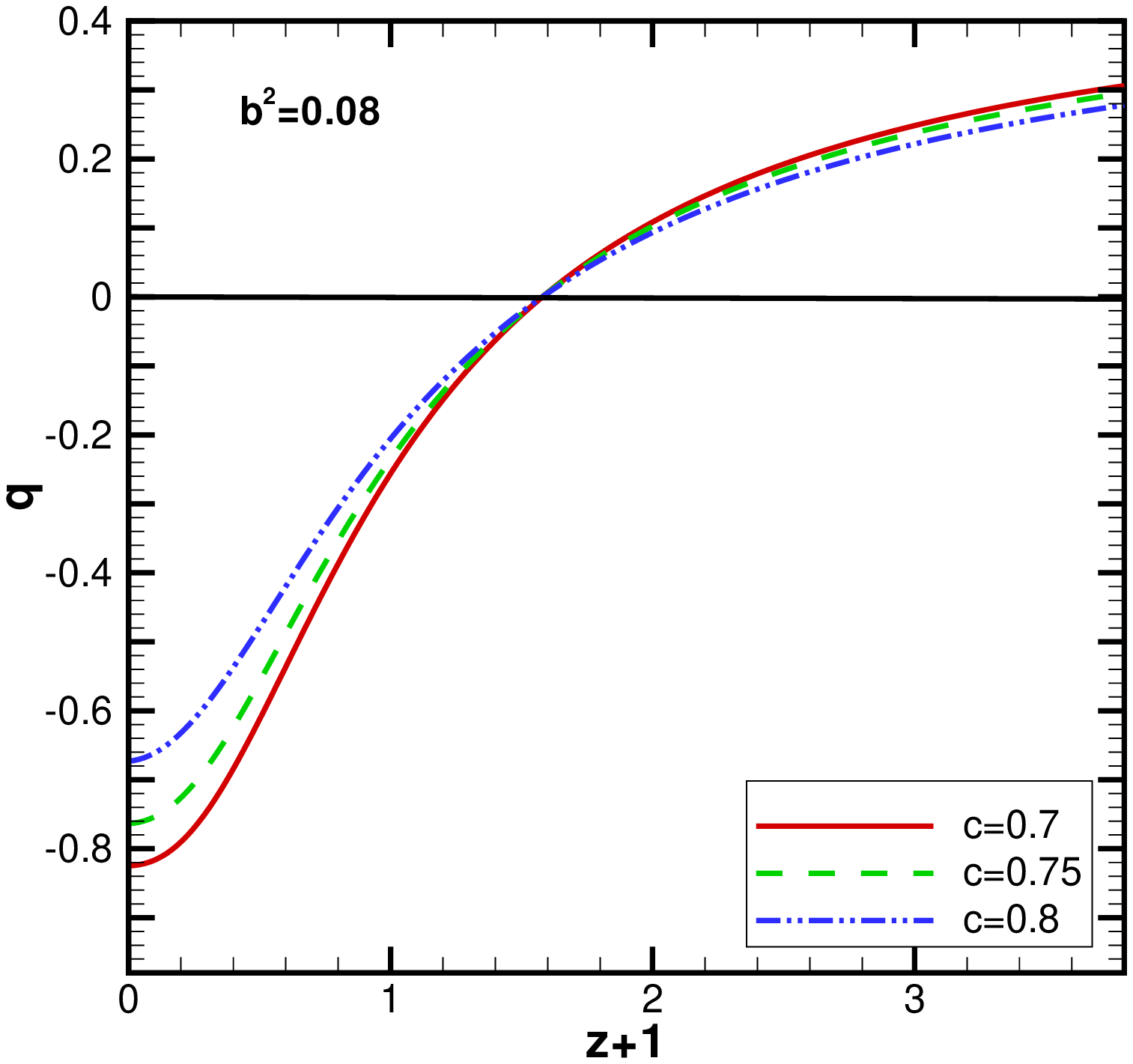}
\includegraphics[width=8cm]{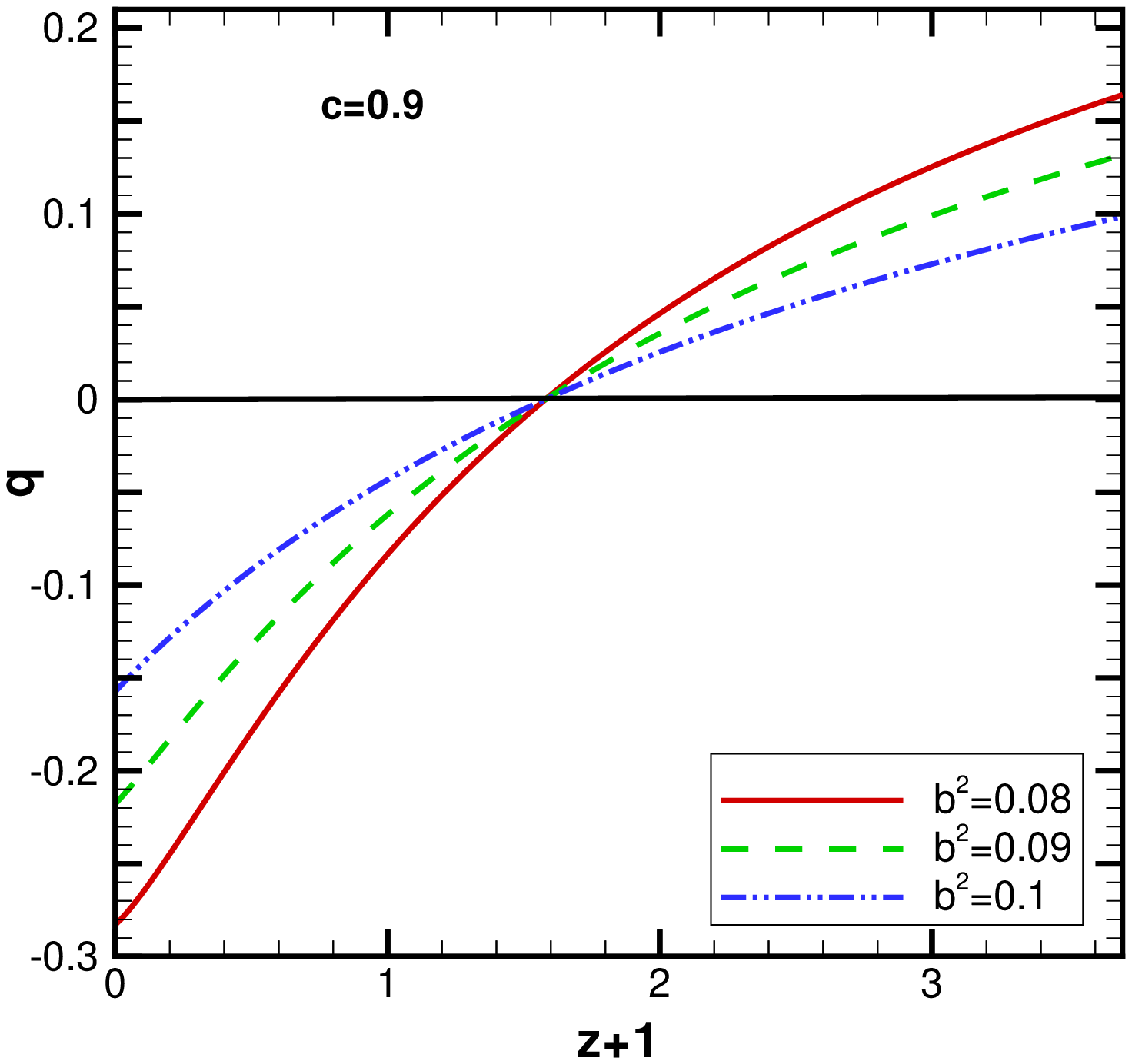}
\caption{The evolution of the deceleration parameter $q$ versus
redshift parameter $z$ for interacting new HDE in DGP
braneworld.}\label{q2}
\end{center}
\end{figure}
Using relation (\ref{OmegaDE1}), one can rewrite Eqs.
(\ref{QWDE3}) and (\ref{Qq3}) in the following form
\begin{eqnarray}\label{QWDE4}
\omega_D&=&\frac{2c^2(c^2-\Omega_D)+\frac{b^2}{c^2}(3\Omega_D-2c^2)(3\Omega_D-c^2)}
{\Omega_D\Big(c^2+\Omega_Dc^2+c^4-3\Omega_D\Big)}, \\
q&=&-\frac{c^4+6\Omega_D-2c^2\Omega_D-5c^2+3b^2(2c^2-3\Omega_D)}
{\Big(c^2+\Omega_Dc^2+c^4-3\Omega_D\Big)}. \label{Qq4}
\end{eqnarray}
In the absence of the interaction, i.e., $b^2=0$, Eqs.
(\ref{QWDE4}) and (\ref{Qq4}) reduce to (\ref{WDEe3}) and
(\ref{qq3}), respectively. Again from relation (\ref{OmegaDE1}) we
see that the limit of standard cosmology where $\Omega_{r_c}$
corresponds to $\Omega_D=c^2$, and in this limit the results of
standard cosmology for interacting HDE with Hubble radius as
systems's IR cutoff is restored \cite{Pav`on},
\begin{eqnarray}\label{St}
\omega_D&=&-\frac{b^2}{c^2(1-c^2)}, \\
q&=&\frac{1}{2}-\frac{3 b^2}{2(1-c^2)}.
\end{eqnarray}
Again, we see that for the choice of $L=H^{-1}$, in standard
cosmology, an interaction is the only way to have an EoS different
from that for dust \cite{Pav`on,Zim}. However, for the model
presented in this paper,  and in the context of DGP braneworld,
even in the absence of interaction, the the natural and simple
choice of $L=H^{-1}$, can leads to the accelerated Universe. This
is one of the main result we have addressed in the present paper.

The evolution of the $\omega_D$ and $q$ are shown in figures
\ref{w2} and \ref{q2} for different model parameters. In one
figure we fix $b^2$ and set different values for $c$, while in
another figure we fix $c$ and set different values for $b^2$. From
these figures we clearly see that the transition from the
deceleration into acceleration phase can occur around $z\sim0.6$,
which is consist with recent observations.
%%%%%%%%%%%%%%%%%%%%%%%%%%%%%%%%%%%%%%%%%%%%%%%%%%%%%%%%%%%%%%%%%%%%%%%%%%
\section{Conclusions and discussions}\label{Con}
The energy density expression of the HDE is based on the area law
of the black hole entropy, and thus any modification of the area
law leads to a modified holographic energy density. Since in the
DGP braneworld the area law of the entropy can be modified as
given in (\ref{Sbh}), thus it is more reasonable to propose a new
energy density for the HDE which is based on this modified entropy
expression.

In this paper, we first showed that the Friedmann equation
describing the evolution of the FRW Universe in DGP  can be recast
as the first law of thermodynamics on the apparent horizon. This
procedure leads to extract an entropy expression associated with
the apparent horizon. We expect the entropy of the black hole
horizon to have the same expression but replacing the apparent
horizon radius $\tilde{r}_A$ with the black hole horizon radius
$r_{+}$. Then, inspired by the entropy expression associated with
the apparent horizon of FRW universe in DGP braneworld, we
proposed a new model for the HDE in the framework of the DGP brane
cosmology. We believe that if one is interested in studying the HDE
in the framework of DGP braneworld, one should take the energy
density in the form of (\ref{rhoD}) instead of the usual form
$\rho_D= {3c^2M_{\rm p}^2}/{L^2}$. Compared to the energy density
of the HDE in standard cosmology, the new HDE describing by
expression (\ref{rhoD}) consists an additional term which
incorporates the effects of un-compact extra dimension onto the
brane. Clearly in the limiting case where the effects of the extra
dimension vanishes,($L\ll3r_{c}$), the energy density of usual HDE
is recovered.

Then, we studied the cosmological implications of this new model
for flat FRW Universe on the brane. For this purpose, we chose the
Hubble radius $L=H^{-1}$, as system IR cutoff. The Hubble radius
is not only the most natural and obvious but also the simplest
choice for IR cutoff in flat universe. It was already shown that
the Hubble radius in flat Universe can result an accelerated
Universe provided the interaction between the two dark components
of the universe is taken into account \cite{Pav`on}. In other
words, in the absence of interaction it leads to the EoS of dust,
$w\omega_D=0$ \cite{Hsu}. Interestingly enough, we found that, for
the new HDE in DGP braneworld, the identification of IR cutoff
with Hubble horizon, $L=H^{-1}$, can lead to an acceleration of
the universe expansion, even in the absence of the interaction
between two dark components. This is contrast to the HDE in
standard cosmology, where $\omega_D=0$ if one choose $L=H^{-1}$ in
the absence of interaction between the two dark components. We
also examined our model by taking into account the interaction
term and derived the cosmological parameters. In order to see the
behavior of the EoS and deceleration parameter, we plotted these
parameters versus redshift in both cases. Our studies show that in
both cases, our Universe has a transition from deceleration to the
acceleration phase around $z\approx0.6$, which is compatible with
observational evidences \cite{Daly,Kom1,Kom2}.
%--------------------------------------------------------------------------------
\acknowledgments{We thank from the Research Council of Shiraz
University. This work has been supported financially by Research
Institute for Astronomy \& Astrophysics of Maragha (RIAAM), Iran.
}%%%%%%%%%%%%%%%%%%%%%%%%%%%%%%%%%%%%%%%%%%%%%%%

\end{document}